\documentclass[aps,prb,reprint,superscriptaddress,showpacs]{revtex4-1}

\usepackage{graphicx} 
\usepackage{subfigure}  
\usepackage{hyperref}   

\begin{document}

\title{Hybrid metal-dielectric ring resonators for optical magnetic metamaterials down to ultraviolet range}

\author{Jianwei Tang}
\affiliation{Centre for Optical and Electromagnetic Research, JORCEP[KTH-ZJU Joint Research Center of Photonics],\\Zhejiang University (ZJU), Hangzhou 310058, China}
\author{Sailing He}
\email[]{sailing@kth.se}
\affiliation{Centre for Optical and Electromagnetic Research, JORCEP[KTH-ZJU Joint Research Center of Photonics],\\Zhejiang University (ZJU), Hangzhou 310058, China}
\affiliation{Department of Electromagnetic Engineering, School of Electrical Engineering,\\Royal Institute of Technology(KTH), 100 44 Stockholm, Sweden}

\date{\today}

\begin{abstract}
In this paper, we derive a model from Maxwell equations for the magnetic resonance of ring resonators. Using this model we revisit the scaling of split-ring resonators. Inspired by our model, we propose a hybrid metal-dielectric ring resonator composed of high index dielectric material (e.g. SiC, TiO$_2$, ZnS) (for the major portion) and metal (e.g. Ag). Such a new magnetic metamaterial is able to overcome the saturation problem of split-ring resonators and therefore is able to operate at short wavelength down to ultraviolet range. 
\end{abstract}


\maketitle

\section{\label{introduction} introduction}
Magnetic response of nature material is usually very weak. With the help of metamaterials, people managed to realize strong magnetic response from microwave frequencies to optical frequencies
\cite{smith2000composite,yen2004terahertz,linden2004magnetic,%
zhang2005midinfrared,enkrich2005magnetic,grigorenko2005nanofabricated,shalaev2005negative,dolling2005cut,zhang2005experimental,%
klein2006single,cai2007metamagnetics,shalaev2007optical,soukoulis2007negative,valentine2008three,chettiar2008optical,%
lahiri2010magnetic,jeyaram2010magnetic,chen2011optical}. 
Optical magnetism can be realized through metallic structures supporting plasmonic modes able to interact with magnetic field, such as split-ring resonators (SRRs)
\cite{linden2004magnetic,zhang2005midinfrared,enkrich2005magnetic,klein2006single,lahiri2010magnetic,%
chen2011optical} and their derivative structures
\cite{grigorenko2005nanofabricated,dolling2005cut,cai2007metamagnetics}.
However, the operation of metallic structures at optical frequencies is limited by the kinetic energy of the electrons in the metal, leading to saturation of the magnetic response when we push the operating frequency deeper into the optical frequency by size scaling \cite{zhou2005saturation}. Although there is a sustained effort to push magnetic response to shorter and shorter wavelength, there is still lack of elegant designs for realization of magnetisms at short wavelengths, such as blue, violet and ultraviolet region. 

In this paper, we first revisit the scaling of the split-ring resonator and analyze the origin of the saturation of the magnetic response. Unlike Ref.~\onlinecite{zhou2005saturation}, where the kinetic inductance $L_e$ is derived from the kinetic energy of electrons, we derive the effective inductance from Maxwell equations and the dispersive permittivity of metal. From the revisit, we find out an effective inductance $L_c$ which is the same as $L_e$ for frequencies far below the plasma frequency of the metal but significantly larger than $L_e$ for visible wavelength rendering scaling to short wavelength even more difficult. Then we propose a hybrid metal-dielectric ring resonator composed of high index dielectric material (e.g. SiC, TiO$_2$, ZnS) (for the major portion) and metal (e.g. Ag). Such a new magnetic metamaterial is able to operate at short wavelength down to ultraviolet range and the operating principle is inspiring for more designs of short wavelength metamaterials.

\section{\label{model} model description}
\begin{figure}[b!]
\includegraphics[width=3in]{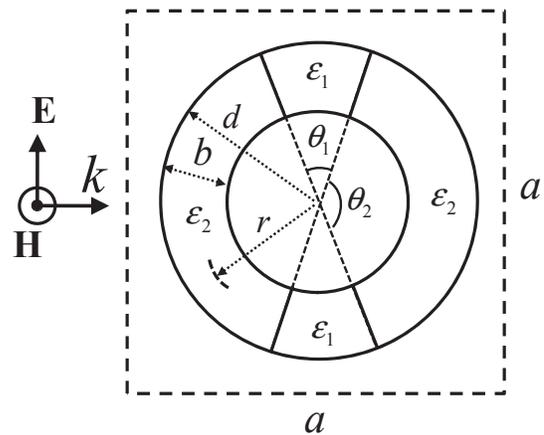}
\caption{\label{fig1}The cross-sectional schematic of the two dimensional ring resonator in a unit cell $a \times a$. The radius $r$ ($r=d-\frac{1}{2}b$) and thickness $b$ of the ring is $r=0.3a$ and $b=0.2a$, respectively. The permittivity of the $2\theta_1$/$2\theta_2$ angle part of the ring is $\epsilon_1$/$\epsilon_2$. The background is air.}
\end{figure}

The first artificial material exhibiting strong magnetic response is a type of ring resonator called split-ring resonator \cite{pendry1999magnetism}, which sees lots of variations, with the key mechanism unchanged. The unit cell ($a \times a$) of the ring resonator metamaterial explored in this paper is shown in Fig.~\ref{fig1}.
It is a two dimensional ring resonator with radius $r=0.3a$ and ring thickness $b=0.2a$. The ring is composed of two types of materials with permittivity $\epsilon_1$ and $\epsilon_2$, occupying $2\theta_1$ and $2\theta_2$ portion by angle, respectively. We assume that $|\epsilon_2|$ is large enough to ensure that the electric field is strongly confined inside the ring so that the electric field inside the ring has no radial component. We also assume a uniform distribution in the radial direction of the electric field inside the ring, so that the displacement current in the ring can be expressed as
\begin{equation} \label{eq:current}
I=b\frac{\partial \mathbf{D}}{\partial t}=-\mathrm{i}\omega b \epsilon_1 E_1=-\mathrm{i}\omega b \epsilon_2 E_2.
\end{equation}
According to Maxwell equations, we have 
\begin{equation} \label{eq:maxwell}
2r\theta_1 E_1+2r\theta_2 E_2=\mathrm{i}\omega(\Phi^{\mathrm{int}}+\Phi^{\mathrm{ext}})
\end{equation}
where $\Phi^{\mathrm{int}}=(1-F)L_\mathrm{g}I$  is the magnetic flux produced by the induced displacement current and $\Phi^{\mathrm{ext}}=\mu_0 H_0 FS$ ($S=\pi a^2$ is the area of the unit cell) is the external driving magnetic flux by incident magnetic field $H_0$. Depolarizing field contributes to the $(1-F)$ factor of $\Phi^{\mathrm{int}}$ where $F=\pi r^2/a^2$ is the filling factor. It accounts for the magnetic coupling between unit cells. Then we derive from Eqs.~\ref{eq:current} and \ref{eq:maxwell} the following equation
\begin{equation} \label{eq:main}
\Bigl[\frac{1}{-\mathrm{i} \omega C}+(1-F)(-\mathrm{i} \omega L_{\mathrm{g}})\Bigr]I=\mathrm{i}\omega \Phi^{\mathrm{ext}}
\end{equation}
where $C=(\frac{1}{C_1}+\frac{1}{C_2})^{-1}$ is the effective capacitance, which is the serial capacitance of $C_1(=(b \epsilon_1)/(2r \theta_1))$ and $C_2(=(b \epsilon_2)/(2r \theta_2))$; $L_\mathrm{g}$ is the effective magnetic inductance of the ring, which can be approximately evaluated as $\mu_0FS$. As long as the real part of $C$ is positive, the ring will exhibit a magnetic resonance when $\omega=1/\sqrt{L_\mathrm{g}C}$. A magnetic resonance is possible for any of the following three cases: (1) $\epsilon_1>0$, $\epsilon_2>0$; (2) $\epsilon_1>0$, $\epsilon_2<0$; (3) $\epsilon_1<0$, $\epsilon_2>0$. The dielectric magnetic metamaterials belong to the first case, which can be applied in microwave or terahertz ranges where large permittivity of the order several tens of $\epsilon_0$ can be easily achieved, but is a challenging issue in the infrared and visible frequency range due to low refractive index of materials in this range \cite{zhao2009mie}. Only recently, optical magnetism in the midinfrared was realized with tellurium dielectric cubic resonators \cite{ginn2012realizing}. Optical magnetism in visible can also be realized using single crystalline silicon \cite{evlyukhin2012demonstration,kuznetsov2012magnetic}, but the fabrication method is quite limited in order to guarantee that the constituent silicon is single crystalline. The conventional metallic SRR at optical wavelength is of the second case, which will be studied in detail in Sec.~\ref{revisit} to revisit the problem of saturation. The third case is a brand new type, which is very intriguing and significant owing to its ability to overcome the saturation problem of conventional SRRs for magnetic response at short wavelength and will be studied in detail in Sec.~\ref{saturation}.

\section{\label{revisit} Scaling of conventional Split-Ring Resonator}
\begin{figure}[h!]
\subfigure{\label{fig2a}
\includegraphics[width=3in]{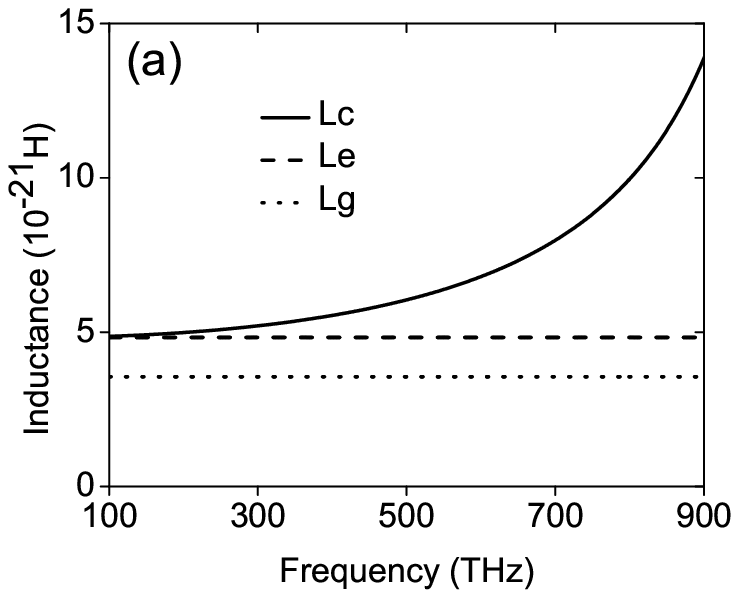}}
\subfigure{\label{fig2b}
\includegraphics[width=3in]{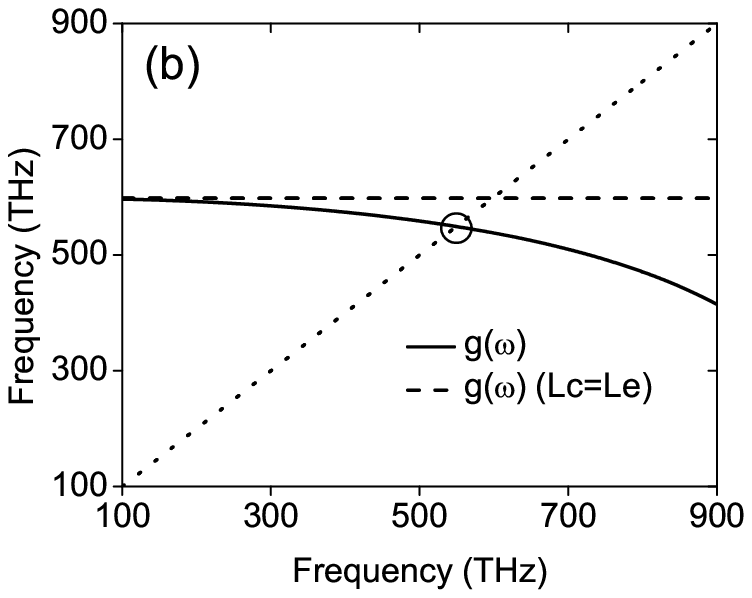}}
\caption{\label{fig2}(a) The comparison between the effective inductance $L_\mathrm{c}$ (solid line), the well-known kinetic inductance $L_\mathrm{e}$ (dash line) and the geometrical inductance $L_\mathrm{g}$ (dot line). (b) Determination of the magnetic resonant frequency (denoted by the circle) with dispersive effective inductance $L_\mathrm{c}$. The structure parameters are: $a=100\mathrm{nm},r=30\mathrm{nm},b=20\mathrm{nm},\theta_1=20^{\circ}.$}
\end{figure}
To analyze the scaling of the conventional SRR at optical wavelength, we use Ag as metal material $\epsilon_2$and air as the dielectric material $\epsilon_1$. The permittivity of Ag can be described by Drude model 
$\epsilon_2=\epsilon_0(\epsilon_\infty-\frac{\omega_\mathrm{p}^2}{\omega^2+\mathrm{i} \gamma \omega})$, which is dispersive and negative below plasma frequency $\omega_\mathrm{p}$. We determined the parameters $\epsilon_\infty(=4)$, $\omega_\mathrm{p}(=1.4\times 10^{16} \mathrm{rad})$ and $\gamma(=9.69 \times 10^{13} \mathrm{rad})$ by fitting to the experimental data \cite{johnson1972optical} in the frequency range 400 $-$ 900 THz. Note that due to the interband transitions, the fitted Drude model is no longer accurate enough above 700 THz (i.e. below 430 nm wavelength), but we will still use the Drude model above 700 THz for principle demonstration. Since $\epsilon_2$ is dispersive, $C_2$ is also dispersive. Because of the negative real part and the imaginary part of $\epsilon_2$, $C_2$ is a negative effective capacitance in parallel connection with a resistance, which is equivalent to a positive inductance $L_c$ in series connection with a resistance $R$, i.e. $(-\mathrm{i}\omega C_2)^{-1}=-\mathrm{i}\omega L_c+R$. Then the effective magnetic permeability can be expressed as a resonant form
\begin{equation} \label{eq:mu_SRR}
\mu_\mathrm{eff}=1-\frac{F'\omega^2}{\omega^2-\frac{1}{(L_\mathrm{g}+L_\mathrm{c})C_1}+\mathrm{i}\frac{\omega R}{L_\mathrm{g}+L_\mathrm{c}}},
\end{equation}
where $F'=\frac{L_\mathrm{g}}{L_\mathrm{g}+L_\mathrm{c}}F$ and $R=\frac{2r\theta_1}{b \epsilon_0 \omega_\mathrm{p}^2}\gamma$. $L_\mathrm{c}$ is dispersive as shown in Fig.~\ref{fig2a} (solid line). The resonant frequency of the effective permeability can be determined by the intersection point of the line   $g(\omega)=1/\sqrt{(L_\mathrm{g}+L_\mathrm{c})C_1}$ (solid line in Fig.~\ref{fig2b}) and the frequency line (dot line in Fig.~\ref{fig2b}). For frequencies far below the plasma frequency of the metal (e.g. below 100 THz), $\epsilon_2$ can be approximated as $\epsilon_2=\epsilon_0(-\frac{\omega_\mathrm{p}^2}{\omega^2+\mathrm{i}\gamma \omega})$. With this approximation, $L_\mathrm{c}$ becomes the well-known kinetic inductance $L_\mathrm{e}=\frac{2r\theta_2}{b\epsilon_0 \omega_\mathrm{p}^2}$ \cite{zhou2005saturation}, which is non-dispersive as shown in Fig.~\ref{fig2a} (dash line). The resonance frequency can then be directly expressed as $1/\sqrt{(L_\mathrm{g}+L_\mathrm{c})C_1}$ (dash line in Fig.~\ref{fig2b}), and the quality factor $Q=R^{-1}\sqrt{(L_\mathrm{g}+L_\mathrm{c})/C_1}$. For the 2D SRR as shown in Fig.~\ref{fig1}, when we scale all the geometrical parameters ($a,b,r$), $L_\mathrm{c}$(equal to $L_\mathrm{e}$), $C_1$ and $R$ keep constant, while $L_\mathrm{g}$ scales proportional to the scale factor as $a^2$. For silver SRR structure operating at optical frequencies, $L_\mathrm{c}$ is comparable with or even larger than $L_\mathrm{g}$, thus reducing $L_\mathrm{g}$ by size scaling is not effective to push resonant wavelength to short optical wavelength. For frequencies not far from the plasma frequency of the metal, $L_\mathrm{c}$ increases quickly when the wavelength decreases (cf. solid line in Fig.~\ref{fig2a}), which makes it even more difficult to push resonant wavelength to shorter optical wavelength by size scaling. We can see from Fig.~\ref{fig2b} by comparing the solid line and dash line that for the given structure parameters, the resonance frequency determined by LC model with the approximation $L_\mathrm{c}=L_\mathrm{e}$ is over-estimated by about 50 THz. There have been efforts to reduce $L_\mathrm{c}$ by using metal with larger plasma frequency, e.g. aluminum \cite{jeyaram2010magnetic}. However, the loss of aluminum is much larger than silver, which means that though magnetic resonance at short wavelength can be achieved, it is very weak. For silver SRR, another problem during scaling is that although the quality factor decrease slowly, $F'$ decrease very quickly, which reduces the amplitude of the magnetic resonance. So the conventional metallic SRRs have their limitations for applications at short optical wavelength. In fact, there are two more factors that further reduce the amplitude of the magnetic resonance at short optical wavelength. In our analysis above, there are two assumptions that are less appropriate when $|\epsilon_2|$ is not large enough (e.g. when approaching the plasma frequency) as compared with the permittivity of the background material. First, the electric field would not be strongly confined inside the ring. Second, apart from the displacement current inside the ring, the displacement current near the ring, whose direction is opposite to the displacement current inside the ring, should also be added to the total current in Eq.~\ref{eq:current}.

\begin{figure}[t]
\includegraphics[width=3.5in]{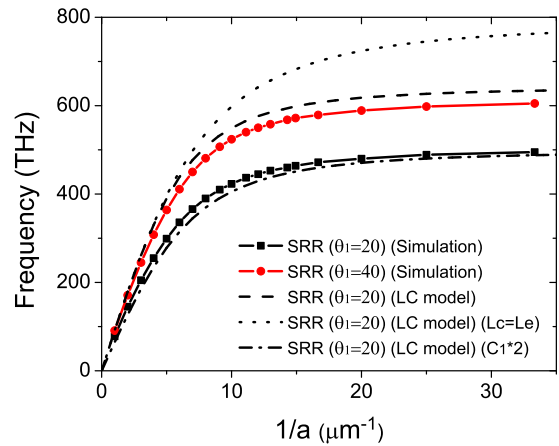}
\caption{\label{fig3}(color online). The scaling of the simulated (solid lines) and LC model calculated (non-solid lines) magnetic resonance frequency of conventional SRRs as a function of the unit cell size $a$.The black and red solid lines are simulated results for SRRs with $\theta_1=20^\circ$ and $40^\circ$, respectively. The dash line is the LC model calculated result for SRRs with  $\theta_1=20^\circ$. The dot line is the LC model calculated result with the approximation $L_\mathrm{c}=L_\mathrm{e}$ for SRRs with $\theta_1=20^\circ$. The dash-dot line is the LC model calculated result with $C_1$ doubled for SRRs with $\theta_1=20^\circ$.}
\end{figure}

In Fig.~\ref{fig3}, we show the scaling process of the SRR structures in term of magnetic resonance frequency, where the solid lines represent simulation results while the dash and dot lines represent LC model results. From the simulation results, we can see the saturation phenomenon when the scaling factor $a$ is small. SRRs with $\theta_1=20^\circ$ saturates at about 500 THz, whereas SRRs with $\theta_1=40^\circ$ saturates at about 600 THz. This is because $\theta_1$ gives smaller $C_1$. We found again in this figure the different results given by LC model with (black dash line) and without (black dot line) the approximation $L_\mathrm{c}=L_\mathrm{e}$. When the scaling factor $1/a$ is small (e.g. $<6\mu\mathrm{m}^{-1}$), the over-estimation is not apparent, which means that in this region dispersive $L_\mathrm{c}$ can be well reduced to non-dispersive $L_\mathrm{e}$. However, when the scaling factor is larger, the over-estimation is quite apparent, up to 130 THz. In this region, we have to use dispersive inductance. We would like to point out that the LC model (black dash line) can well describe the saturation phenomenon, although there is a frequency shift compared to the simulation result (black solid line). Such a frequency shift between our LC model and simulation is not surprising. We have assumed a uniform distribution in the radial direction of the electric field inside the ring to get Eq.~\ref{eq:current}, whereas the numerical simulation shows that the electric field is far from uniform. This affects the estimation of the effective capacitance $C_1$ in our model, as has been evidenced by Ref.~\onlinecite{zhou2005saturation}. The frequency shift is larger for larger $\theta_1$. Keeping this in mind, for SRRs with $\theta_1=20^\circ$ we can simply double our effective capacitance $L_\mathrm{c}$ to well reproduce (black dash-dot line) the simulation result. In fact, we do not intend a very accurate theoretic model which requires lots of detailed factors to be involved at the cost of losing simplicity and physical intuitiveness to some degree. Our LC model aims at revealing the resonance conditions and key influential factors thus inspiring new designs with specific features of performance, e.g. magnetic metamaterial for operation at short wavelength as described in Sec.~\ref{saturation}.        

\section{\label{saturation} Hybrid metal-dielectric ring resonator to overcome the saturation problem}

\begin{figure}[t]
\subfigure{\label{fig4a}
\includegraphics[width=3.5in]{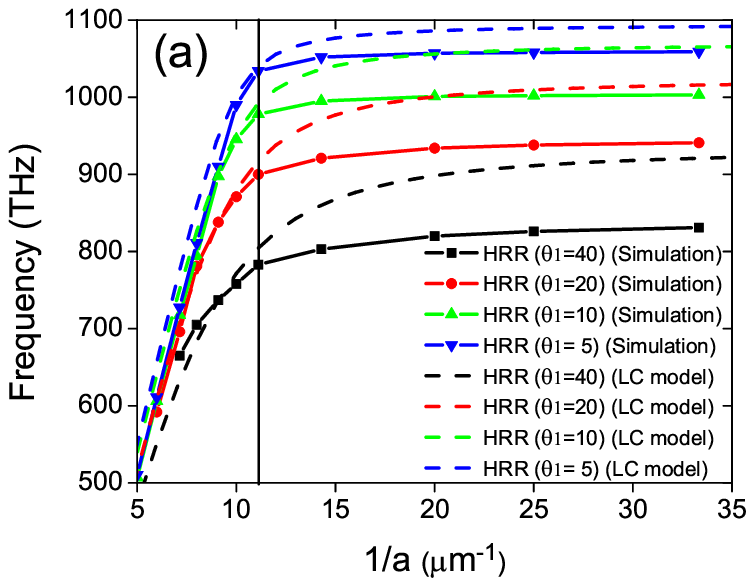}}
\subfigure{\label{fig4b}
\includegraphics[width=3.5in]{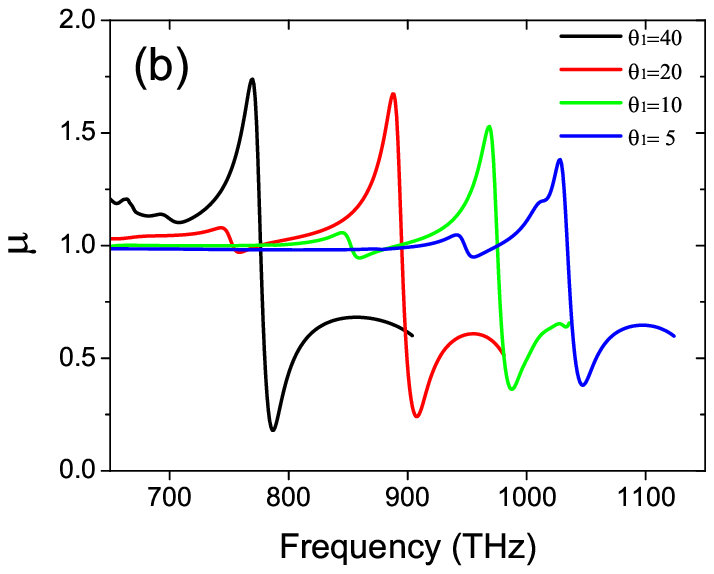}}
\caption{\label{fig4}(Color online). (a) The scaling of the simulated (solid lines) and LC model calculated (dash lines) magnetic resonance frequency of HRRs as a function of the unit cell size $a$; (b) The effective permeability (retrieved from simulated S parameters) of HRRs with $a=90\mathrm{nm}$ (denoted by the vertical line in panel (a)) and $\theta_1=40^\circ,20^\circ,10^\circ,5^\circ$, from left to right.}
\end{figure}

\begin{figure}[b]
\includegraphics[width=3.5in]{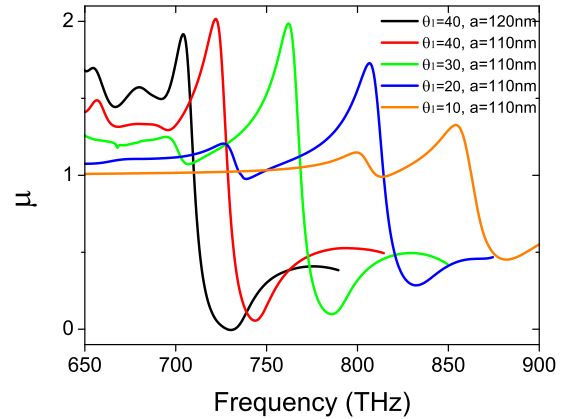}
\caption{\label{fig5}(Color online). The simulated effective permeability of HRR using experimental material data of Ag from literature.}
\end{figure}

In order to operate at short wavelength, we proposed a new type of ring resonator, of which
$\epsilon_2$ is high index dielectric material while $\epsilon_1$ is metal. This is the case $\epsilon_1<0$, $\epsilon_2>0$. We refer to this new type of ring resonator as HRR (Hybrid Ring Resonator) for short in this paper. For HRR, 
$\epsilon_1=\epsilon_0(\epsilon_\infty-\frac{\omega_\mathrm{p}^2}{\omega^2+\mathrm{i} \gamma \omega})$. For demonstration, we assume $\epsilon_2=6\epsilon_0$, which is reasonable since low-loss dielectric materials with permittivity around this value at short wavelength are available by using e.g. TiO$_2$, SiC or ZnS from various deposition technologies. The effective capacitance $C_2=(b\epsilon_2)/(2r\theta_2)$ is positive while $C_2=(b\epsilon_1)/(2r\theta_1)$ is negative and is equivalent to a positive inductance $L_\mathrm{c}$ in series connection with a resistance $R$. The effective permeability has the same form as Eq.~\ref{eq:mu_SRR}
\begin{equation} \label{eq:mu_DSRR}
\mu_\mathrm{eff}=1-\frac{F'\omega^2}{\omega^2-\frac{1}{(L_\mathrm{g}+L_\mathrm{c})C_2}+\mathrm{i}\frac{\omega R}{L_\mathrm{g}+L_\mathrm{c}}},
\end{equation}
where $F'=\frac{L_\mathrm{g}}{L_\mathrm{g}+L_\mathrm{c}}F$ and $R=\frac{2r\theta_1}{b \epsilon_0
\omega_\mathrm{p}^2}\gamma$. For frequencies far below $\omega_\mathrm{p}$, we can approximate $\epsilon_1$ as $\epsilon_0(-\frac{\omega_\mathrm{p}^2}{\omega^2+\mathrm{i}\gamma \omega})$ so that $L_\mathrm{c}=\frac{2r\theta_1}{b\epsilon_0 \omega_\mathrm{p}^2}$ and the resonant frequency can be expressed as $1/\sqrt{(L_\mathrm{g}+L_\mathrm{c})C_2}$. We can now analyze why HRR is able to operate at short wavelength with strong magnetic response. First, $L_\mathrm{c}$ is proportional to the length of the metal part, i.e. $\theta_1$, hence $L_\mathrm{c}$ of HRR is much smaller than that of the conventional SRR and we can thus expect a much shorter resonant wavelength. Second, we can expect a higher quality factor of resonance because the resistance $R$ is also proportional to $\theta_1$. Third, a smaller $L_\mathrm{c}$ also gives a larger $F'$, thus a larger amplitude of magnetic resonance. Additionally, unlike conventional SRR, HRR doesn't become dysfunctional when the permittivity of the metal is too small to confine electric field near plasma frequency. This is because the high index dielectric material plays the dominant role in confining the electric filed inside the ring and the displacement current near the ring has the same direction as the displacement current inside the ring.

The scaling of HRR is represented in Fig.~\ref{fig4a} for $\theta_1=40^\circ$ (black lines), $20^\circ$ (red lines), $10^\circ$ (green lines) and $5^\circ$ (blue lines) by using LC model (dash lines) and simulation (solid lines) results. For each scaling curve (i.e. constant $\theta_1$), we still observe the saturation phenomenon as expected by our LC model. However, the saturation frequencies are much larger than those of conventional SRRs (cf. Fig.~\ref{fig3}) mainly owing to the reduction of $L_\mathrm{c}$. If we consider $\theta_1$ as a variable parameter, we can see that the HRR is able to overcome the saturation problem. To demonstrate this ability, we keep $a$ constant as 90 nm (as indicated by the vertical line in Fig.~\ref{fig4a}) and decrease $\theta_1$ from $40^\circ$ to $5^\circ$. The effective permeability spectra are plotted in Fig.~\ref{fig4b}. By changing $\theta_1$, the resonance frequency increase from 800 THz to 1050 THz. It's worthwhile to point out that according to our Drude model of Ag, the real part of the permittivity is only -0.5 at 1050 THz and is positive at larger than 1115 THz. We can almost say that we managed to push magnetic resonance frequency to as high as the permittivity of metal is negative. Additionally, as expected by our analysis with LC model, the amplitude and quality factor of the magnetic resonances keep quite healthy, only decreasing gently while being pushed to higher frequency.  

Till now we have simulated HRRs using Drude model in order to compare with the LC model and conventional SRRs. However, we know that due to the interband transitions, the fitted Drude model is no longer accurate above 700 THz. Therefore, we check again the performance of HRR by simulation using experimental material data of Ag \cite{johnson1972optical}. There are two main differences between the experimental data and the Drude model. First, the frequency point where the permittivity change from negative to positive is about 920 THz for the experimental data, which is much lower than Drude model (1115 THz); Second, the loss is several times higher at frequencies larger than 800 THz for the experimental data. The simulated effective permeability spectra are shown in Fig.~\ref{fig5}. The magnetic resonance can still be pushed to frequency close to 920 THz, although the amplitude and quality factor are not good as the case using Drude model Ag due to higher loss for the experimental Ag above 800 THz.

\section{\label{conclusion} conclusion}
We have proposed a model for ring resonators, based on which we summarized three cases of magnetic resonances and studied in detail two cases in optical frequency. In our LC model, we utilized the effective kinetic inductance $L_\mathrm{c}$, which is more accurate than $L_\mathrm{e}$ in describing the performance of ring resonators, especially at short wavelength. In the framework of our model, we revisited the scaling of conventional SRRs and proposed a hybrid metal-dielectric ring resonator which is able to overcome the saturation problem of SRRs hence a very elegant design to operate at short wavelength, down to ultraviolet wavelength. Our model is physically intuitive to guide magnetic metamaterial design in optical frequency. Our proposal of hybrid ring resonator design is just a prototype, variations and improvements are possible for better performance and ease of fabrication.   

\begin{acknowledgments}
This work is partially supported by National High Technology Research and Development Program (863 Program)
of China (No. 2012AA030402), the National Natural Science Foundation of China (Nos. 60990322 and 61178062), Swedish VR grant (No. 621-2011-4620) and AOARD (114045).
\end{acknowledgments}

\bibliography{apstemplate}

\begin{thebibliography}{25}%
\makeatletter
\providecommand \@ifxundefined [1]{%
 \@ifx{#1\undefined}
}%
\providecommand \@ifnum [1]{%
 \ifnum #1\expandafter \@firstoftwo
 \else \expandafter \@secondoftwo
 \fi
}%
\providecommand \@ifx [1]{%
 \ifx #1\expandafter \@firstoftwo
 \else \expandafter \@secondoftwo
 \fi
}%
\providecommand \natexlab [1]{#1}%
\providecommand \enquote  [1]{``#1''}%
\providecommand \bibnamefont  [1]{#1}%
\providecommand \bibfnamefont [1]{#1}%
\providecommand \citenamefont [1]{#1}%
\providecommand \href@noop [0]{\@secondoftwo}%
\providecommand \href [0]{\begingroup \@sanitize@url \@href}%
\providecommand \@href[1]{\@@startlink{#1}\@@href}%
\providecommand \@@href[1]{\endgroup#1\@@endlink}%
\providecommand \@sanitize@url [0]{\catcode `\\12\catcode `\$12\catcode
  `\&12\catcode `\#12\catcode `\^12\catcode `\_12\catcode `\%12\relax}%
\providecommand \@@startlink[1]{}%
\providecommand \@@endlink[0]{}%
\providecommand \url  [0]{\begingroup\@sanitize@url \@url }%
\providecommand \@url [1]{\endgroup\@href {#1}{\urlprefix }}%
\providecommand \urlprefix  [0]{URL }%
\providecommand \Eprint [0]{\href }%
\providecommand \doibase [0]{http://dx.doi.org/}%
\providecommand \selectlanguage [0]{\@gobble}%
\providecommand \bibinfo  [0]{\@secondoftwo}%
\providecommand \bibfield  [0]{\@secondoftwo}%
\providecommand \translation [1]{[#1]}%
\providecommand \BibitemOpen [0]{}%
\providecommand \bibitemStop [0]{}%
\providecommand \bibitemNoStop [0]{.\EOS\space}%
\providecommand \EOS [0]{\spacefactor3000\relax}%
\providecommand \BibitemShut  [1]{\csname bibitem#1\endcsname}%
\let\auto@bib@innerbib\@empty
\bibitem [{\citenamefont {Smith}\ \emph {et~al.}(2000)\citenamefont {Smith},
  \citenamefont {Padilla}, \citenamefont {Vier}, \citenamefont {Nemat-Nasser},\
  and\ \citenamefont {Schultz}}]{smith2000composite}%
  \BibitemOpen
  \bibfield  {author} {\bibinfo {author} {\bibfnamefont {D.}~\bibnamefont
  {Smith}}, \bibinfo {author} {\bibfnamefont {W.}~\bibnamefont {Padilla}},
  \bibinfo {author} {\bibfnamefont {D.}~\bibnamefont {Vier}}, \bibinfo {author}
  {\bibfnamefont {S.}~\bibnamefont {Nemat-Nasser}}, \ and\ \bibinfo {author}
  {\bibfnamefont {S.}~\bibnamefont {Schultz}},\ }\href@noop {} {\bibfield
  {journal} {\bibinfo  {journal} {Physical review letters}\ }\textbf {\bibinfo
  {volume} {84}},\ \bibinfo {pages} {4184} (\bibinfo {year}
  {2000})}\BibitemShut {NoStop}%
\bibitem [{\citenamefont {Yen}\ \emph {et~al.}(2004)\citenamefont {Yen},
  \citenamefont {Padilla}, \citenamefont {Fang}, \citenamefont {Vier},
  \citenamefont {Smith}, \citenamefont {Pendry}, \citenamefont {Basov},\ and\
  \citenamefont {Zhang}}]{yen2004terahertz}%
  \BibitemOpen
  \bibfield  {author} {\bibinfo {author} {\bibfnamefont {T.}~\bibnamefont
  {Yen}}, \bibinfo {author} {\bibfnamefont {W.}~\bibnamefont {Padilla}},
  \bibinfo {author} {\bibfnamefont {N.}~\bibnamefont {Fang}}, \bibinfo {author}
  {\bibfnamefont {D.}~\bibnamefont {Vier}}, \bibinfo {author} {\bibfnamefont
  {D.}~\bibnamefont {Smith}}, \bibinfo {author} {\bibfnamefont
  {J.}~\bibnamefont {Pendry}}, \bibinfo {author} {\bibfnamefont
  {D.}~\bibnamefont {Basov}}, \ and\ \bibinfo {author} {\bibfnamefont
  {X.}~\bibnamefont {Zhang}},\ }\href@noop {} {\bibfield  {journal} {\bibinfo
  {journal} {Science}\ }\textbf {\bibinfo {volume} {303}},\ \bibinfo {pages}
  {1494} (\bibinfo {year} {2004})}\BibitemShut {NoStop}%
\bibitem [{\citenamefont {Linden}\ \emph {et~al.}(2004)\citenamefont {Linden},
  \citenamefont {Enkrich}, \citenamefont {Wegener}, \citenamefont {Zhou},
  \citenamefont {Koschny},\ and\ \citenamefont
  {Soukoulis}}]{linden2004magnetic}%
  \BibitemOpen
  \bibfield  {author} {\bibinfo {author} {\bibfnamefont {S.}~\bibnamefont
  {Linden}}, \bibinfo {author} {\bibfnamefont {C.}~\bibnamefont {Enkrich}},
  \bibinfo {author} {\bibfnamefont {M.}~\bibnamefont {Wegener}}, \bibinfo
  {author} {\bibfnamefont {J.}~\bibnamefont {Zhou}}, \bibinfo {author}
  {\bibfnamefont {T.}~\bibnamefont {Koschny}}, \ and\ \bibinfo {author}
  {\bibfnamefont {C.}~\bibnamefont {Soukoulis}},\ }\href@noop {} {\bibfield
  {journal} {\bibinfo  {journal} {Science}\ }\textbf {\bibinfo {volume}
  {306}},\ \bibinfo {pages} {1351} (\bibinfo {year} {2004})}\BibitemShut
  {NoStop}%
\bibitem [{\citenamefont {Zhang}\ \emph
  {et~al.}(2005{\natexlab{a}})\citenamefont {Zhang}, \citenamefont {Fan},
  \citenamefont {Minhas}, \citenamefont {Frauenglass}, \citenamefont {Malloy},\
  and\ \citenamefont {Brueck}}]{zhang2005midinfrared}%
  \BibitemOpen
  \bibfield  {author} {\bibinfo {author} {\bibfnamefont {S.}~\bibnamefont
  {Zhang}}, \bibinfo {author} {\bibfnamefont {W.}~\bibnamefont {Fan}}, \bibinfo
  {author} {\bibfnamefont {B.}~\bibnamefont {Minhas}}, \bibinfo {author}
  {\bibfnamefont {A.}~\bibnamefont {Frauenglass}}, \bibinfo {author}
  {\bibfnamefont {K.}~\bibnamefont {Malloy}}, \ and\ \bibinfo {author}
  {\bibfnamefont {S.}~\bibnamefont {Brueck}},\ }\href@noop {} {\bibfield
  {journal} {\bibinfo  {journal} {Physical review letters}\ }\textbf {\bibinfo
  {volume} {94}},\ \bibinfo {pages} {37402} (\bibinfo {year}
  {2005}{\natexlab{a}})}\BibitemShut {NoStop}%
\bibitem [{\citenamefont {Enkrich}\ \emph {et~al.}(2005)\citenamefont
  {Enkrich}, \citenamefont {Wegener}, \citenamefont {Linden}, \citenamefont
  {Burger}, \citenamefont {Zschiedrich}, \citenamefont {Schmidt}, \citenamefont
  {Zhou}, \citenamefont {Koschny},\ and\ \citenamefont
  {Soukoulis}}]{enkrich2005magnetic}%
  \BibitemOpen
  \bibfield  {author} {\bibinfo {author} {\bibfnamefont {C.}~\bibnamefont
  {Enkrich}}, \bibinfo {author} {\bibfnamefont {M.}~\bibnamefont {Wegener}},
  \bibinfo {author} {\bibfnamefont {S.}~\bibnamefont {Linden}}, \bibinfo
  {author} {\bibfnamefont {S.}~\bibnamefont {Burger}}, \bibinfo {author}
  {\bibfnamefont {L.}~\bibnamefont {Zschiedrich}}, \bibinfo {author}
  {\bibfnamefont {F.}~\bibnamefont {Schmidt}}, \bibinfo {author} {\bibfnamefont
  {J.}~\bibnamefont {Zhou}}, \bibinfo {author} {\bibfnamefont {T.}~\bibnamefont
  {Koschny}}, \ and\ \bibinfo {author} {\bibfnamefont {C.}~\bibnamefont
  {Soukoulis}},\ }\href@noop {} {\bibfield  {journal} {\bibinfo  {journal}
  {Physical review letters}\ }\textbf {\bibinfo {volume} {95}},\ \bibinfo
  {pages} {203901} (\bibinfo {year} {2005})}\BibitemShut {NoStop}%
\bibitem [{\citenamefont {Grigorenko}\ \emph {et~al.}(2005)\citenamefont
  {Grigorenko}, \citenamefont {Geim}, \citenamefont {Gleeson}, \citenamefont
  {Zhang}, \citenamefont {Firsov}, \citenamefont {Khrushchev},\ and\
  \citenamefont {Petrovic}}]{grigorenko2005nanofabricated}%
  \BibitemOpen
  \bibfield  {author} {\bibinfo {author} {\bibfnamefont {A.}~\bibnamefont
  {Grigorenko}}, \bibinfo {author} {\bibfnamefont {A.}~\bibnamefont {Geim}},
  \bibinfo {author} {\bibfnamefont {H.}~\bibnamefont {Gleeson}}, \bibinfo
  {author} {\bibfnamefont {Y.}~\bibnamefont {Zhang}}, \bibinfo {author}
  {\bibfnamefont {A.}~\bibnamefont {Firsov}}, \bibinfo {author} {\bibfnamefont
  {I.}~\bibnamefont {Khrushchev}}, \ and\ \bibinfo {author} {\bibfnamefont
  {J.}~\bibnamefont {Petrovic}},\ }\href@noop {} {\bibfield  {journal}
  {\bibinfo  {journal} {Nature}\ }\textbf {\bibinfo {volume} {438}},\ \bibinfo
  {pages} {335} (\bibinfo {year} {2005})}\BibitemShut {NoStop}%
\bibitem [{\citenamefont {Shalaev}\ \emph {et~al.}(2005)\citenamefont
  {Shalaev}, \citenamefont {Cai}, \citenamefont {Chettiar}, \citenamefont
  {Yuan}, \citenamefont {Sarychev}, \citenamefont {Drachev},\ and\
  \citenamefont {Kildishev}}]{shalaev2005negative}%
  \BibitemOpen
  \bibfield  {author} {\bibinfo {author} {\bibfnamefont {V.}~\bibnamefont
  {Shalaev}}, \bibinfo {author} {\bibfnamefont {W.}~\bibnamefont {Cai}},
  \bibinfo {author} {\bibfnamefont {U.}~\bibnamefont {Chettiar}}, \bibinfo
  {author} {\bibfnamefont {H.}~\bibnamefont {Yuan}}, \bibinfo {author}
  {\bibfnamefont {A.}~\bibnamefont {Sarychev}}, \bibinfo {author}
  {\bibfnamefont {V.}~\bibnamefont {Drachev}}, \ and\ \bibinfo {author}
  {\bibfnamefont {A.}~\bibnamefont {Kildishev}},\ }\href@noop {} {\bibfield
  {journal} {\bibinfo  {journal} {Optics Letters}\ }\textbf {\bibinfo {volume}
  {30}},\ \bibinfo {pages} {3356} (\bibinfo {year} {2005})}\BibitemShut
  {NoStop}%
\bibitem [{\citenamefont {Dolling}\ \emph {et~al.}(2005)\citenamefont
  {Dolling}, \citenamefont {Enkrich}, \citenamefont {Wegener}, \citenamefont
  {Zhou}, \citenamefont {Soukoulis},\ and\ \citenamefont
  {Linden}}]{dolling2005cut}%
  \BibitemOpen
  \bibfield  {author} {\bibinfo {author} {\bibfnamefont {G.}~\bibnamefont
  {Dolling}}, \bibinfo {author} {\bibfnamefont {C.}~\bibnamefont {Enkrich}},
  \bibinfo {author} {\bibfnamefont {M.}~\bibnamefont {Wegener}}, \bibinfo
  {author} {\bibfnamefont {J.}~\bibnamefont {Zhou}}, \bibinfo {author}
  {\bibfnamefont {C.}~\bibnamefont {Soukoulis}}, \ and\ \bibinfo {author}
  {\bibfnamefont {S.}~\bibnamefont {Linden}},\ }\href@noop {} {\bibfield
  {journal} {\bibinfo  {journal} {Optics letters}\ }\textbf {\bibinfo {volume}
  {30}},\ \bibinfo {pages} {3198} (\bibinfo {year} {2005})}\BibitemShut
  {NoStop}%
\bibitem [{\citenamefont {Zhang}\ \emph
  {et~al.}(2005{\natexlab{b}})\citenamefont {Zhang}, \citenamefont {Fan},
  \citenamefont {Panoiu}, \citenamefont {Malloy}, \citenamefont {Osgood},\ and\
  \citenamefont {Brueck}}]{zhang2005experimental}%
  \BibitemOpen
  \bibfield  {author} {\bibinfo {author} {\bibfnamefont {S.}~\bibnamefont
  {Zhang}}, \bibinfo {author} {\bibfnamefont {W.}~\bibnamefont {Fan}}, \bibinfo
  {author} {\bibfnamefont {N.}~\bibnamefont {Panoiu}}, \bibinfo {author}
  {\bibfnamefont {K.}~\bibnamefont {Malloy}}, \bibinfo {author} {\bibfnamefont
  {R.}~\bibnamefont {Osgood}}, \ and\ \bibinfo {author} {\bibfnamefont
  {S.}~\bibnamefont {Brueck}},\ }\href@noop {} {\bibfield  {journal} {\bibinfo
  {journal} {Physical review letters}\ }\textbf {\bibinfo {volume} {95}},\
  \bibinfo {pages} {137404} (\bibinfo {year} {2005}{\natexlab{b}})}\BibitemShut
  {NoStop}%
\bibitem [{\citenamefont {Klein}\ \emph {et~al.}(2006)\citenamefont {Klein},
  \citenamefont {Enkrich}, \citenamefont {Wegener}, \citenamefont {Soukoulis},\
  and\ \citenamefont {Linden}}]{klein2006single}%
  \BibitemOpen
  \bibfield  {author} {\bibinfo {author} {\bibfnamefont {M.}~\bibnamefont
  {Klein}}, \bibinfo {author} {\bibfnamefont {C.}~\bibnamefont {Enkrich}},
  \bibinfo {author} {\bibfnamefont {M.}~\bibnamefont {Wegener}}, \bibinfo
  {author} {\bibfnamefont {C.}~\bibnamefont {Soukoulis}}, \ and\ \bibinfo
  {author} {\bibfnamefont {S.}~\bibnamefont {Linden}},\ }\href@noop {}
  {\bibfield  {journal} {\bibinfo  {journal} {Optics letters}\ }\textbf
  {\bibinfo {volume} {31}},\ \bibinfo {pages} {1259} (\bibinfo {year}
  {2006})}\BibitemShut {NoStop}%
\bibitem [{\citenamefont {Cai}\ \emph {et~al.}(2007)\citenamefont {Cai},
  \citenamefont {Chettiar}, \citenamefont {Yuan}, \citenamefont {De~Silva},
  \citenamefont {Kildishev}, \citenamefont {Drachev},\ and\ \citenamefont
  {Shalaev}}]{cai2007metamagnetics}%
  \BibitemOpen
  \bibfield  {author} {\bibinfo {author} {\bibfnamefont {W.}~\bibnamefont
  {Cai}}, \bibinfo {author} {\bibfnamefont {U.}~\bibnamefont {Chettiar}},
  \bibinfo {author} {\bibfnamefont {H.}~\bibnamefont {Yuan}}, \bibinfo {author}
  {\bibfnamefont {V.}~\bibnamefont {De~Silva}}, \bibinfo {author}
  {\bibfnamefont {A.}~\bibnamefont {Kildishev}}, \bibinfo {author}
  {\bibfnamefont {V.}~\bibnamefont {Drachev}}, \ and\ \bibinfo {author}
  {\bibfnamefont {V.}~\bibnamefont {Shalaev}},\ }\href@noop {} {\bibfield
  {journal} {\bibinfo  {journal} {Optics Express}\ }\textbf {\bibinfo {volume}
  {15}},\ \bibinfo {pages} {3333} (\bibinfo {year} {2007})}\BibitemShut
  {NoStop}%
\bibitem [{\citenamefont {Shalaev}(2007)}]{shalaev2007optical}%
  \BibitemOpen
  \bibfield  {author} {\bibinfo {author} {\bibfnamefont {V.}~\bibnamefont
  {Shalaev}},\ }\href@noop {} {\bibfield  {journal} {\bibinfo  {journal}
  {Nature photonics}\ }\textbf {\bibinfo {volume} {1}},\ \bibinfo {pages} {41}
  (\bibinfo {year} {2007})}\BibitemShut {NoStop}%
\bibitem [{\citenamefont {Soukoulis}\ \emph {et~al.}(2007)\citenamefont
  {Soukoulis}, \citenamefont {Linden},\ and\ \citenamefont
  {Wegener}}]{soukoulis2007negative}%
  \BibitemOpen
  \bibfield  {author} {\bibinfo {author} {\bibfnamefont {C.}~\bibnamefont
  {Soukoulis}}, \bibinfo {author} {\bibfnamefont {S.}~\bibnamefont {Linden}}, \
  and\ \bibinfo {author} {\bibfnamefont {M.}~\bibnamefont {Wegener}},\
  }\href@noop {} {\bibfield  {journal} {\bibinfo  {journal} {Science}\ }\textbf
  {\bibinfo {volume} {315}},\ \bibinfo {pages} {47} (\bibinfo {year}
  {2007})}\BibitemShut {NoStop}%
\bibitem [{\citenamefont {Valentine}\ \emph {et~al.}(2008)\citenamefont
  {Valentine}, \citenamefont {Zhang}, \citenamefont {Zentgraf}, \citenamefont
  {Ulin-Avila}, \citenamefont {Genov}, \citenamefont {Bartal},\ and\
  \citenamefont {Zhang}}]{valentine2008three}%
  \BibitemOpen
  \bibfield  {author} {\bibinfo {author} {\bibfnamefont {J.}~\bibnamefont
  {Valentine}}, \bibinfo {author} {\bibfnamefont {S.}~\bibnamefont {Zhang}},
  \bibinfo {author} {\bibfnamefont {T.}~\bibnamefont {Zentgraf}}, \bibinfo
  {author} {\bibfnamefont {E.}~\bibnamefont {Ulin-Avila}}, \bibinfo {author}
  {\bibfnamefont {D.}~\bibnamefont {Genov}}, \bibinfo {author} {\bibfnamefont
  {G.}~\bibnamefont {Bartal}}, \ and\ \bibinfo {author} {\bibfnamefont
  {X.}~\bibnamefont {Zhang}},\ }\href@noop {} {\bibfield  {journal} {\bibinfo
  {journal} {Nature}\ }\textbf {\bibinfo {volume} {455}},\ \bibinfo {pages}
  {376} (\bibinfo {year} {2008})}\BibitemShut {NoStop}%
\bibitem [{\citenamefont {Chettiar}\ \emph {et~al.}(2008)\citenamefont
  {Chettiar}, \citenamefont {Xiao}, \citenamefont {Kildishev}, \citenamefont
  {Cai}, \citenamefont {Yuan}, \citenamefont {Drachev},\ and\ \citenamefont
  {Shalaev}}]{chettiar2008optical}%
  \BibitemOpen
  \bibfield  {author} {\bibinfo {author} {\bibfnamefont {U.}~\bibnamefont
  {Chettiar}}, \bibinfo {author} {\bibfnamefont {S.}~\bibnamefont {Xiao}},
  \bibinfo {author} {\bibfnamefont {A.}~\bibnamefont {Kildishev}}, \bibinfo
  {author} {\bibfnamefont {W.}~\bibnamefont {Cai}}, \bibinfo {author}
  {\bibfnamefont {H.}~\bibnamefont {Yuan}}, \bibinfo {author} {\bibfnamefont
  {V.}~\bibnamefont {Drachev}}, \ and\ \bibinfo {author} {\bibfnamefont
  {V.}~\bibnamefont {Shalaev}},\ }\href@noop {} {\bibfield  {journal} {\bibinfo
   {journal} {MRS bulletin}\ }\textbf {\bibinfo {volume} {33}},\ \bibinfo
  {pages} {921} (\bibinfo {year} {2008})}\BibitemShut {NoStop}%
\bibitem [{\citenamefont {Lahiri}\ \emph {et~al.}(2010)\citenamefont {Lahiri},
  \citenamefont {McMeekin}, \citenamefont {Khokhar}, \citenamefont
  {De~La~Rue},\ and\ \citenamefont {Johnson}}]{lahiri2010magnetic}%
  \BibitemOpen
  \bibfield  {author} {\bibinfo {author} {\bibfnamefont {B.}~\bibnamefont
  {Lahiri}}, \bibinfo {author} {\bibfnamefont {S.}~\bibnamefont {McMeekin}},
  \bibinfo {author} {\bibfnamefont {A.}~\bibnamefont {Khokhar}}, \bibinfo
  {author} {\bibfnamefont {R.}~\bibnamefont {De~La~Rue}}, \ and\ \bibinfo
  {author} {\bibfnamefont {N.}~\bibnamefont {Johnson}},\ }\href@noop {}
  {\bibfield  {journal} {\bibinfo  {journal} {Optics Express}\ }\textbf
  {\bibinfo {volume} {18}},\ \bibinfo {pages} {3210} (\bibinfo {year}
  {2010})}\BibitemShut {NoStop}%
\bibitem [{\citenamefont {Jeyaram}\ \emph {et~al.}(2010)\citenamefont
  {Jeyaram}, \citenamefont {Jha}, \citenamefont {Agio}, \citenamefont
  {L{\"o}ffler},\ and\ \citenamefont {Ekinci}}]{jeyaram2010magnetic}%
  \BibitemOpen
  \bibfield  {author} {\bibinfo {author} {\bibfnamefont {Y.}~\bibnamefont
  {Jeyaram}}, \bibinfo {author} {\bibfnamefont {S.}~\bibnamefont {Jha}},
  \bibinfo {author} {\bibfnamefont {M.}~\bibnamefont {Agio}}, \bibinfo {author}
  {\bibfnamefont {J.}~\bibnamefont {L{\"o}ffler}}, \ and\ \bibinfo {author}
  {\bibfnamefont {Y.}~\bibnamefont {Ekinci}},\ }\href@noop {} {\bibfield
  {journal} {\bibinfo  {journal} {Optics letters}\ }\textbf {\bibinfo {volume}
  {35}},\ \bibinfo {pages} {1656} (\bibinfo {year} {2010})}\BibitemShut
  {NoStop}%
\bibitem [{\citenamefont {Chen}\ \emph {et~al.}(2011)\citenamefont {Chen},
  \citenamefont {Chen}, \citenamefont {Wu}, \citenamefont {Sun}, \citenamefont
  {Zhou}, \citenamefont {Guo}, \citenamefont {Hsiao}, \citenamefont {Yang},
  \citenamefont {Zheludev},\ and\ \citenamefont {Tsai}}]{chen2011optical}%
  \BibitemOpen
  \bibfield  {author} {\bibinfo {author} {\bibfnamefont {W.}~\bibnamefont
  {Chen}}, \bibinfo {author} {\bibfnamefont {C.}~\bibnamefont {Chen}}, \bibinfo
  {author} {\bibfnamefont {P.}~\bibnamefont {Wu}}, \bibinfo {author}
  {\bibfnamefont {S.}~\bibnamefont {Sun}}, \bibinfo {author} {\bibfnamefont
  {L.}~\bibnamefont {Zhou}}, \bibinfo {author} {\bibfnamefont {G.}~\bibnamefont
  {Guo}}, \bibinfo {author} {\bibfnamefont {C.}~\bibnamefont {Hsiao}}, \bibinfo
  {author} {\bibfnamefont {K.}~\bibnamefont {Yang}}, \bibinfo {author}
  {\bibfnamefont {N.}~\bibnamefont {Zheludev}}, \ and\ \bibinfo {author}
  {\bibfnamefont {D.}~\bibnamefont {Tsai}},\ }\href@noop {} {\bibfield
  {journal} {\bibinfo  {journal} {Optics Express}\ }\textbf {\bibinfo {volume}
  {19}},\ \bibinfo {pages} {12837} (\bibinfo {year} {2011})}\BibitemShut
  {NoStop}%
\bibitem [{\citenamefont {Zhou}\ \emph {et~al.}(2005)\citenamefont {Zhou},
  \citenamefont {Koschny}, \citenamefont {Kafesaki}, \citenamefont {Economou},
  \citenamefont {Pendry},\ and\ \citenamefont
  {Soukoulis}}]{zhou2005saturation}%
  \BibitemOpen
  \bibfield  {author} {\bibinfo {author} {\bibfnamefont {J.}~\bibnamefont
  {Zhou}}, \bibinfo {author} {\bibfnamefont {T.}~\bibnamefont {Koschny}},
  \bibinfo {author} {\bibfnamefont {M.}~\bibnamefont {Kafesaki}}, \bibinfo
  {author} {\bibfnamefont {E.}~\bibnamefont {Economou}}, \bibinfo {author}
  {\bibfnamefont {J.}~\bibnamefont {Pendry}}, \ and\ \bibinfo {author}
  {\bibfnamefont {C.}~\bibnamefont {Soukoulis}},\ }\href@noop {} {\bibfield
  {journal} {\bibinfo  {journal} {Physical review letters}\ }\textbf {\bibinfo
  {volume} {95}},\ \bibinfo {pages} {223902} (\bibinfo {year}
  {2005})}\BibitemShut {NoStop}%
\bibitem [{\citenamefont {Pendry}\ \emph {et~al.}(1999)\citenamefont {Pendry},
  \citenamefont {Holden}, \citenamefont {Robbins},\ and\ \citenamefont
  {Stewart}}]{pendry1999magnetism}%
  \BibitemOpen
  \bibfield  {author} {\bibinfo {author} {\bibfnamefont {J.}~\bibnamefont
  {Pendry}}, \bibinfo {author} {\bibfnamefont {A.}~\bibnamefont {Holden}},
  \bibinfo {author} {\bibfnamefont {D.}~\bibnamefont {Robbins}}, \ and\
  \bibinfo {author} {\bibfnamefont {W.}~\bibnamefont {Stewart}},\ }\href@noop
  {} {\bibfield  {journal} {\bibinfo  {journal} {Microwave Theory and
  Techniques, IEEE Transactions on}\ }\textbf {\bibinfo {volume} {47}},\
  \bibinfo {pages} {2075} (\bibinfo {year} {1999})}\BibitemShut {NoStop}%
\bibitem [{\citenamefont {Zhao}\ \emph {et~al.}(2009)\citenamefont {Zhao},
  \citenamefont {Zhou}, \citenamefont {Zhang},\ and\ \citenamefont
  {Lippens}}]{zhao2009mie}%
  \BibitemOpen
  \bibfield  {author} {\bibinfo {author} {\bibfnamefont {Q.}~\bibnamefont
  {Zhao}}, \bibinfo {author} {\bibfnamefont {J.}~\bibnamefont {Zhou}}, \bibinfo
  {author} {\bibfnamefont {F.}~\bibnamefont {Zhang}}, \ and\ \bibinfo {author}
  {\bibfnamefont {D.}~\bibnamefont {Lippens}},\ }\href@noop {} {\bibfield
  {journal} {\bibinfo  {journal} {Materials Today}\ }\textbf {\bibinfo {volume}
  {12}},\ \bibinfo {pages} {60} (\bibinfo {year} {2009})}\BibitemShut {NoStop}%
\bibitem [{\citenamefont {Ginn}\ \emph {et~al.}(2012)\citenamefont {Ginn},
  \citenamefont {Brener}, \citenamefont {Peters}, \citenamefont {Wendt},
  \citenamefont {Stevens}, \citenamefont {Hines}, \citenamefont {Basilio},
  \citenamefont {Warne}, \citenamefont {Ihlefeld}, \citenamefont {Clem} \emph
  {et~al.}}]{ginn2012realizing}%
  \BibitemOpen
  \bibfield  {author} {\bibinfo {author} {\bibfnamefont {J.}~\bibnamefont
  {Ginn}}, \bibinfo {author} {\bibfnamefont {I.}~\bibnamefont {Brener}},
  \bibinfo {author} {\bibfnamefont {D.}~\bibnamefont {Peters}}, \bibinfo
  {author} {\bibfnamefont {J.}~\bibnamefont {Wendt}}, \bibinfo {author}
  {\bibfnamefont {J.}~\bibnamefont {Stevens}}, \bibinfo {author} {\bibfnamefont
  {P.}~\bibnamefont {Hines}}, \bibinfo {author} {\bibfnamefont
  {L.}~\bibnamefont {Basilio}}, \bibinfo {author} {\bibfnamefont
  {L.}~\bibnamefont {Warne}}, \bibinfo {author} {\bibfnamefont
  {J.}~\bibnamefont {Ihlefeld}}, \bibinfo {author} {\bibfnamefont
  {P.}~\bibnamefont {Clem}},  \emph {et~al.},\ }\href@noop {} {\bibfield
  {journal} {\bibinfo  {journal} {Physical Review Letters}\ }\textbf {\bibinfo
  {volume} {108}},\ \bibinfo {pages} {97402} (\bibinfo {year}
  {2012})}\BibitemShut {NoStop}%
\bibitem [{\citenamefont {Evlyukhin}\ \emph {et~al.}(2012)\citenamefont
  {Evlyukhin}, \citenamefont {Novikov}, \citenamefont {Zywietz}, \citenamefont
  {Eriksen}, \citenamefont {Reinhardt}, \citenamefont {Bozhevolnyi},\ and\
  \citenamefont {Chichkov}}]{evlyukhin2012demonstration}%
  \BibitemOpen
  \bibfield  {author} {\bibinfo {author} {\bibfnamefont {A.}~\bibnamefont
  {Evlyukhin}}, \bibinfo {author} {\bibfnamefont {S.}~\bibnamefont {Novikov}},
  \bibinfo {author} {\bibfnamefont {U.}~\bibnamefont {Zywietz}}, \bibinfo
  {author} {\bibfnamefont {R.}~\bibnamefont {Eriksen}}, \bibinfo {author}
  {\bibfnamefont {C.}~\bibnamefont {Reinhardt}}, \bibinfo {author}
  {\bibfnamefont {S.}~\bibnamefont {Bozhevolnyi}}, \ and\ \bibinfo {author}
  {\bibfnamefont {B.}~\bibnamefont {Chichkov}},\ }\href@noop {} {\bibfield
  {journal} {\bibinfo  {journal} {Nano Letters}\ }\textbf {\bibinfo {volume}
  {12}},\ \bibinfo {pages} {3749} (\bibinfo {year} {2012})}\BibitemShut
  {NoStop}%
\bibitem [{\citenamefont {Kuznetsov}\ \emph {et~al.}(2012)\citenamefont
  {Kuznetsov}, \citenamefont {Miroshnichenko}, \citenamefont {Fu},
  \citenamefont {Zhang},\ and\ \citenamefont
  {Luk¡¯yanchuk}}]{kuznetsov2012magnetic}%
  \BibitemOpen
  \bibfield  {author} {\bibinfo {author} {\bibfnamefont {A.}~\bibnamefont
  {Kuznetsov}}, \bibinfo {author} {\bibfnamefont {A.}~\bibnamefont
  {Miroshnichenko}}, \bibinfo {author} {\bibfnamefont {Y.}~\bibnamefont {Fu}},
  \bibinfo {author} {\bibfnamefont {J.}~\bibnamefont {Zhang}}, \ and\ \bibinfo
  {author} {\bibfnamefont {B.}~\bibnamefont {Luk¡¯yanchuk}},\ }\href@noop {}
  {\bibfield  {journal} {\bibinfo  {journal} {Scientific Reports}\ }\textbf
  {\bibinfo {volume} {2}},\ \bibinfo {pages} {492} (\bibinfo {year}
  {2012})}\BibitemShut {NoStop}%
\bibitem [{\citenamefont {Johnson}\ and\ \citenamefont
  {Christy}(1972)}]{johnson1972optical}%
  \BibitemOpen
  \bibfield  {author} {\bibinfo {author} {\bibfnamefont {P.}~\bibnamefont
  {Johnson}}\ and\ \bibinfo {author} {\bibfnamefont {R.}~\bibnamefont
  {Christy}},\ }\href@noop {} {\bibfield  {journal} {\bibinfo  {journal}
  {Physical Review B}\ }\textbf {\bibinfo {volume} {6}},\ \bibinfo {pages}
  {4370} (\bibinfo {year} {1972})}\BibitemShut {NoStop}%
\end{thebibliography}%

\end{document}